\documentclass[acmlarge]{acmart}

\usepackage{xeCJK}
\usepackage{amssymb}   
\usepackage{pifont}    

\usepackage{xcolor}
\definecolor{darkgreen}{RGB}{0,100,0}

\newif\ifshowcomments
\showcommentstrue

\ifshowcomments
  \newcommand{\commentyp}[1]{{\color{blue}[{\bf Yulu:} #1]}}
  \newcommand{\notejs}[1]{{\color{orange}[{\bf Jat:} #1]}}
 
    \newcommand{\todo}[1]{{\color{red}[{\bf TODO:} #1]}}
\else
  \newcommand{\commentyp}[1]{}
  \newcommand{\notejs}[1]{}

    \newcommand{\todo}[1]{}
\fi

\newcommand{\changeyp}[1]{{\color{black}#1}}
\newcommand{\changejs}[1]{{\color{black}#1}}

\makeatletter
\newcommand{\myconfshort}{\acmConference@shortname}
\newcommand{\myconffull}{\acmConference@name}
\newcommand{\myconfdate}{\acmConference@date}
\newcommand{\myconfloc}{\acmConference@venue}
\AtBeginDocument{
  \fancypagestyle{firstpagestyle}{
    \fancyhead{}%
    \fancyfoot[C]{}%
  }
  \fancyhf{}
  \fancyhead[LO]{\@headfootfont\shorttitle}%
  \fancyhead[RE]{\@headfootfont\@shortauthors}%
  \fancyhead[LE]{\@headfootfont\footnotesize \myconfshort, \myconfdate, \myconfloc}%
  \fancyhead[RO]{\@headfootfont\footnotesize \myconfshort, \myconfdate, \myconfloc}%
  \fancyfoot[C]{}%
}
\makeatother
\acmBooktitle{\conffull\@ (\confshort), \confdate, \confloc}

\AtBeginDocument{%
  }

\usepackage{float}

\usepackage{tabularx}
\usepackage{subcaption}
\usepackage{caption}
\begin{document}

\copyrightyear{2026}
\acmYear{2026}
\setcopyright{cc}
\setcctype{by}
\acmConference[FAccT '26]{The 2026 ACM Conference on Fairness, Accountability, and Transparency}{June 25--28, 2026}{Montreal, QC, Canada}
\acmBooktitle{The 2026 ACM Conference on Fairness, Accountability, and Transparency (FAccT '26), June 25--28, 2026, Montreal, QC, Canada}
\acmDOI{10.1145/3805689.3806742}
\acmISBN{979-8-4007-2596-8/2026/06}

\title[Understanding the Role of Algorithm Registers in AI Governance]{Understanding the Role of Algorithm Registers in AI Governance Through Comparative Analysis of China and the UK }

\author{Yulu Pi}
\email{yulu.pi@uni-due.de}
\affiliation{%
  \institution{Research Centre Trust, UA Ruhr, University of Duisburg-Essen}
  \city{Duisburg}
  \country{Germany}
}

\author{Wenlong Li}
\email{wenlong.li@zju.edu.cn}
\affiliation{%
  \institution{Guanghua School of Law, Zhejiang University}
  \city{Hangzhou}
  \country{China}
}

\author{Jatinder Singh}
\email{jat@compacctsys.net}
\affiliation{%
  \institution{Research Centre Trust, UA Ruhr, University of Duisburg-Essen}
  \city{Duisburg}
  \country{Germany}
}
\affiliation{%
  \institution{University of Cambridge}
  \city{Cambridge}
  \country{United Kingdom}
}

\renewcommand{\shortauthors}{Pi et al.}
\begin{abstract}

Algorithm registers are increasingly being both considered and deployed as instruments in AI governance. They are often expected to deliver transparency; however, in practice their design, scope, and implementation vary substantially. Currently, we lack a \textit{holistic understanding} of the potential roles that registers might play in AI governance, and how different design choices both shape and reflect those roles. This paper therefore asks \textit{how do algorithm registers differ across jurisdictions, and what do these differences reveal about their roles in AI governance?} Towards this, we conduct a comparative analysis of two influential but contrasting algorithm registration mechanisms, China's Beian system and the UK's Algorithmic Transparency Recording Standard (ATRS), drawing on publicly available regulatory documents, registration guidelines, and registry data. \changeyp{Crucially, our analysis shows that an algorithm register, depending on its design and implementation, can serve functions beyond transparency, including pre-market approval, enabling ecosystem-level understanding, and acting as a broader regulatory infrastructure.} As algorithm registries proliferate globally, we stress the importance of {researchers and policymakers} considering and examining the concrete governance functions that algorithm registries can perform as a result of their design and institutional context, rather than approaching them primarily through a transparency lens.




\end{abstract}

\maketitle

\section{Introduction}

Regulators worldwide are developing and employing governance mechanisms to manage the risks and opportunities presented by algorithmic systems. Among emerging regulatory approaches, \textit{algorithm registers}---broadly understood as databases where organizations document and disclose information about their developed or deployed algorithmic system---have gained global popularity. Many jurisdictions, including China, the Netherlands (NL), the United Kingdom (UK), and several regions in the United States (US), have established their own algorithm registration frameworks~\cite{Penicaud2025}. 

Importantly, algorithmic registers vary in their design, scope and implementation. Some registers are implemented as publicly searchable databases (e.g., the UK and NL), while others rely on limited public disclosure and are primarily accessible to regulators (e.g., China). Disclosure requirements range from using generic templates applied across algorithm types (e.g., the UK and NL) to requiring more detailed, algorithm-specific technical assessments (e.g., China). These differences shape the roles algorithm registers play in AI governance -- what kinds of oversight they enable, for whom, and with what implications. Existing studies have examined algorithm registers in specific national or regional contexts, including China~\cite{Zhang2024PromisePerils,sheehan2024tracing} and Europe~\cite{Box-Ticking,Cath_Jansen2022,algorithmregister2025,AIasserive}, but they are largely described in their local context, non-comparative and 
often tend to adopt a conceptual approach rather than analyzing actual registry data~\cite{Box-Ticking}. Moreover, much of the literature approaches algorithmic registers primarily through a transparency lens, with limited consideration of their wider governance functions. 

As such, we \textbf{currently lack more holistic understandings of the potential roles AI registers can play in AI governance} and how different design choices shape these roles, especially given substantial variation across jurisdictions. This gap limits our ability to understand the full range of functions that algorithm registers can perform. {As the global adoption of AI registers grows, addressing this gap is becoming increasingly critical to ensure they are conceptualised and understood in a more holistic way, as well as built and used more effectively. }




To this end, we ask: \textit{how do algorithm registers differ across jurisdictions, and what do these differences reveal about their roles in AI governance?} We address this question through a systematic comparative analysis of algorithm registers in China and the UK. China and the UK were selected because both countries have implemented mandatory, relatively mature algorithm registration regimes (in contrast to voluntary systems, such as in NL) and hold influential positions in global AI governance. By examining publicly accessible policy documents, registration guidelines, and registry data, our study shows the differences between China's and the UK's algorithm registration mechanisms, demonstrating how these variations reveal the distinct functions these registers serve in AI governance.

Importantly, we do not assess the effectiveness of algorithmic registers, nor do we prescribe how they should function. Evaluating effectiveness first requires a clear and holistic understanding of the functions that registers can perform, how those functions vary across jurisdictions, and which design choices shape them in practice. What counts as an effective register depends on the specific domain it addresses, the regulatory goals at stake, and the ways those goals are implemented and evaluated. These factors can vary substantially across contexts. Rather than presupposing a fixed standard against which registers should be assessed, this paper instead examines how variations in register design and implementation can support different governance aims and outcomes.

By doing so, this paper deepens our understanding of algorithm registers by offering comparative insights that situates them within broader AI governance patterns and lays the groundwork for more context-sensitive evaluations in future research. Overall, we show that algorithm registers can be a flexible mechanism whose functions depends on how they are designed and embedded within regulatory contexts, cautioning against assuming any fixed functions (typically 
transparency) without attention to their concrete design and regulatory context. \changeyp{As algorithm registers gain global prominence, understanding these varied functions is essential for informing and \changeyp{scrutinising} the continued development of algorithm registration systems and ensuring that they serve their intended regulatory purposes.}


\section{Background: Algorithm Registers in AI governance}
\label{sec:background}
An \textit{algorithm register} (also called an algorithm registry or algorithm filing system), is a governance mechanism that is increasingly being adopted worldwide. Countries and jurisdictions implementing such systems include China, the UK, France, Germany, the Netherlands~\cite{Netherlands2022_AlgorithmsRegister}, and several cities in the US, with Canada recently announcing the launch of its first federal AI register in November 2025~\cite{Canada2025_AIRegister}. The terminology and implementation of algorithm registers differ by country. Considering the two countries that are the focus of this study, in the UK the register operates under the Algorithmic Transparency Recording Standard (ATRS), whereas in China the practice is referred to as 备案 (Beian), meaning algorithmic registration and filing. The form of documentation also varies: some registers take the form of web-based searchable databases (e.g., UK), while others consist of static downloadable documents (e.g., China). While its specific focus may vary across countries, in general terms an AI register provides a standardized and archivable way to document algorithmic systems~\cite{PublicAIRegisters2020, vanVliet2024}. 

\changejs{Research on algorithmic registries largely focuses on Europe, the UK, and the US, and typically considers them as tools purporting to enhance transparency,}
which in turn is expected to support trust, oversight by the public and civil society~\cite{omdia_ai_transparency_registers_2021,amnesty_automated_racism_2025, healthcareaireg,Box-Ticking,AIasserive,Artyushina2024}. Proponents of registers argue that by systematically recording the existence, purpose, and operational scope of algorithms, registers make them visible and accessible. This increased visibility allows stakeholders to scrutinize them, identify potential harms or biases, and hold developers and institutions accountable. Yet whether algorithmic registers meaningfully deliver on these aspirations is 
contested~\cite{Box-Ticking,Cath_Jansen2022}. Prior work shows that transparency alone does not guarantee meaningful oversight~\cite{Reviewable,TinFAccT,Miricrim,NullCompliance} and can even foster unwarranted \changejs{reassurance and} trust~\cite{Seeingwithoutknowing,IThinkIGetYourPoint,Placebo}. Critics further argue that many registers are largely performative, risking the production of `transparency theater': governance practices that perform openness without enabling meaningful understanding or accountability~\cite{Cath_Jansen2022,Theaters}. Entries are often limited to high-level descriptions, making it difficult for individuals or regulators to meaningfully assess or challenge algorithmic decisions; some even describe these initiatives as ``futile''~\cite{SheehanDu2022}. 

However, the governance effects of algorithm registers extend beyond what is recorded and made visible. Registers are not only repositories of data but also represent and entail governance processes that shape behavior. The practices of preparing, reviewing, and maintaining records themselves reflect and reconfigure institutional processes, shaping how actors engage with the system and how compliance and accountability is enacted (see~\cite{shreyarecordkeeping}). The process of registration, including the preparation of information by registrants, the review and verification procedures undertaken by regulators, and mechanisms for updating and enforcement, therefore gives rise to different governance effects depending on the particulars of the arrangements.

Importantly, this paper \textit{does not consider whether algorithmic registers deliver meaningful transparency nor to take a stance on their overall effectiveness}. Instead, we contribute empirically by examining how registers are actually designed and implemented across jurisdictions, so to advance a more holistic understanding of their governance functions. By analyzing how registers are designed and operationalized, including the procedures for submission, verification, updating, and enforcement, as well as their institutional embedding, we show how these processes shape what registers can do in practice and inform how they are interpreted and deployed in governance contexts. \changeyp{By grounding our analysis in concrete design choices and institutional processes, we provide a richer basis for evaluating and improving registration mechanisms across different regulatory contexts. This is particularly timely as algorithm registers \changejs{gain attention and} proliferate globally\changejs{, where} policymakers face practical decisions about how to design and implement them: understanding the governance functions that different design choices enable or constrain is a necessary precondition for making relevant decisions.}


\section{Methodology}


This study conducts a structured comparative analysis of algorithm registers in China and the UK to answer the central question: \textit{how do algorithm registers differ across jurisdictions, and what do these differences reveal about their roles in AI governance?} 
To do so, we combine (1) qualitative document analysis of policy texts and implementation materials and (2) quantitative analysis of publicly disclosed register entries. Qualitative document analysis provides insight into the stated objectives of the registers, the requirements they impose, and the guidance offered to meet those requirements. Quantitative analysis of register entries reveals what information is actually recorded and uncovers patterns emerging from these data. Together, these methods enable us to link variations in register design and implementation to their potential governance functions.

China and UK were selected for several reasons. First, they represent very different legal, political, and regulatory contexts in which algorithm registries are embedded~\cite{almaamari2025innovationoversightcrossregionalstudy,roberts2024Global}. Both China and the UK have implemented early, relatively mature, and policy-relevant versions of algorithm registration. Unlike other registers in Europe, the UK's ATRS has been made mandatory across central government, making it more comparable to China's compulsory registration system. Second, both countries exert significant influence on regional and global debates about AI governance, making their approaches important for understanding broader regulatory trajectories. Finally, the study benefits from the authors' complementary institutional positions within research organisations based in both China and the UK, with one researcher drawing on prior industry experience in China. This combination enables an empirically informed and culturally attuned comparative analysis of these two countries.  


\paragraph{Qualitative Document Analysis}
The qualitative component draws on three categories of sources, encompassing all publicly available documents related to algorithm registers obtained through official government channels:

\begin{enumerate}
    \item \textbf{Official Policy/Legal Text:} Formal documents establishing the objectives and requirements of the registers. 
    

    \item \textbf{Supporting Communications and Interpretive Materials.} FAQs, explanatory notes, regulatory announcements, and reports related to the registers issued by relevant authorities.
    
    
    \item \textbf{Operational Guidance and Submission Materials.} Templates, checklists, instructions, and other technical guidance used for completing register submissions.
\end{enumerate}

All documents were obtained from official government websites, regulatory agencies, and public archives. A complete list of the included documents is provided in Appendix \ref{tab:official-documents}


\paragraph{Analytical Framework and Units of Analysis for Qualitative Document Analysis}
\label{analytical_framework}

To conduct a systematic comparison, we adapt and extend the comparative framework for general AI regulatory policy proposed by Roberts et al. (2023)~\cite{RobertsEtAl2023AIFramework}. This general framework is particularly well suited to our study because it captures both high-level regulatory objectives and the operational mechanisms through which regulations are implemented. Based on an initial review of existing algorithm registration practices, we adjust the framework to address the specific context of algorithm registers by adding register-specific dimensions, such as Information Requirements and Disclosure and Maintenance Procedures. The following are the dimensions through which we analyze and compare algorithm registers


\begin{enumerate}
    \item \textbf{Definition and Key Aims}: How the algorithm register is defined and what it is intended to achieve.
    \item \textbf{Legal Basis}: The legal instruments underpinning each system.
    \item \textbf{Scope and Coverage}: Which actors and systems must register, and under what conditions.
    \item \textbf{Information Requirements}: The fields, disclosures, and technical details required for each entry.
    \item \textbf{Disclosure and Maintenance}: How registration information is published, maintained, and updated.
\end{enumerate}

These dimensions help capture the full regulatory lifecycle, from regulatory objective to operationalization, and therefore allow us to analyze not only what each jurisdiction seeks to do but how the register's design enables or constrains those goals. Our focus is therefore on the broader institutional context and operational processes surrounding registers, rather than purely on the information they record. Treating these dimensions as the unit of analysis allows a direct link between regulatory intention and regulatory practice. 


    


\paragraph{Analysis of Register Data}
To complement our document analysis, we conduct a descriptive analysis of publicly accessible register entries. In China, publicly disclosed information is limited to basic information, such as the developer and the stated purpose of the algorithm. We examine the categories of information that are publicly available and contrast them with the broader set of fields required for regulatory approval but not disclosed. Despite these limitations, the data still allows us to track changes in registration over time, identify the types of algorithms registered, and quantify the number of registrations per entity. In the United Kingdom, publicly disclosed register data are more extensive. In addition to conducting analyses comparable to those for China, UK register entries include organizational disclosures on impact assessments and risk-mitigation measures, enabling a deeper examination of registration practices.



\section{Comparative Analysis of Algorithm Registers in China and the UK}
\label{sec:compare}

\begin{figure*}[ht]
    \centering
    \includegraphics[width=0.7\textwidth]{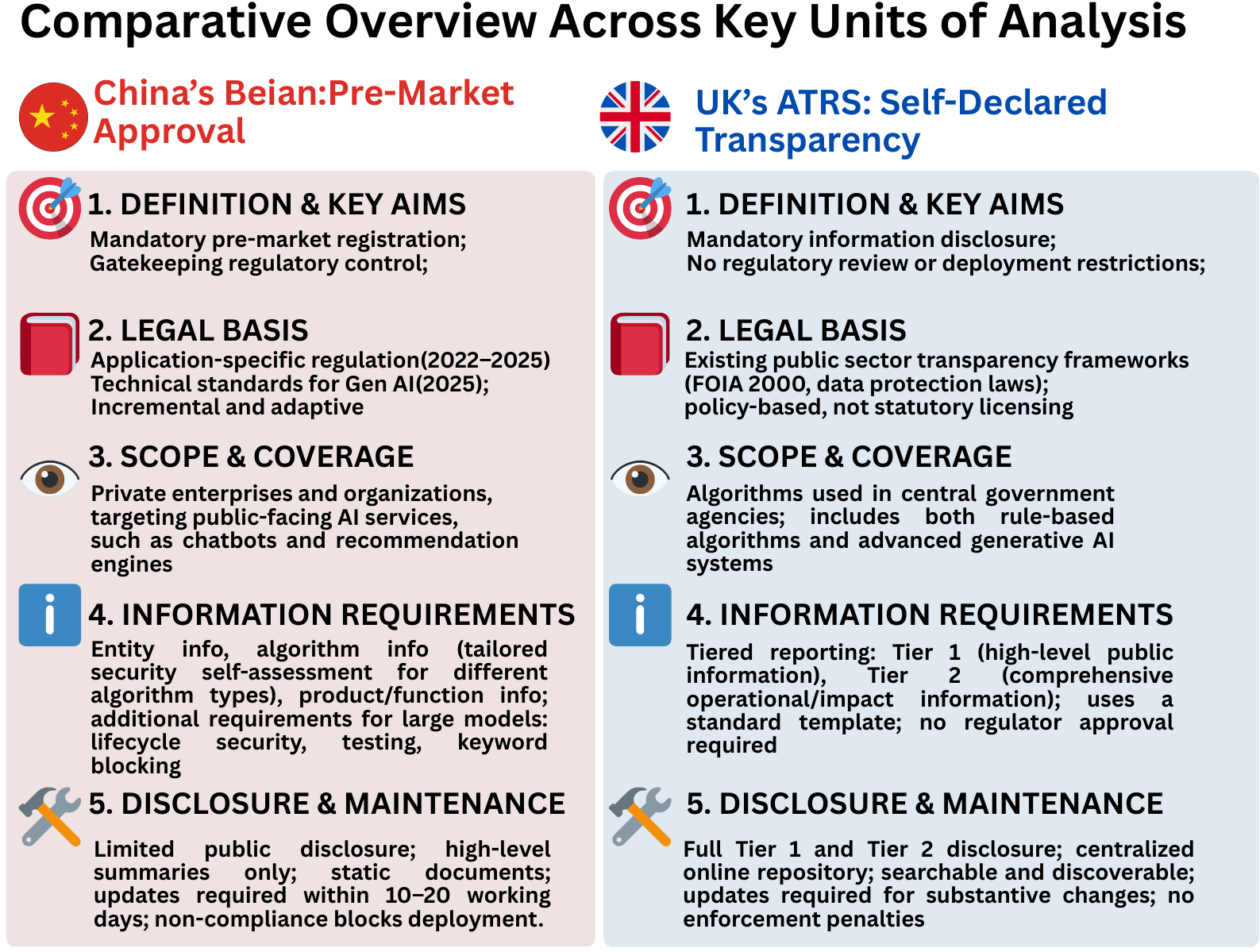}
    \caption{Comparison of China's and the UK's algorithm registration systems across our analytical dimensions.}
    \label{fig:register_comparison}
\end{figure*}

We systematically compare China's and the UK's algorithm registers, following the analytical framework and units of analysis just mentioned (\S\ref{analytical_framework}). Figure~\ref{fig:register_comparison} provides a summarized overview of how the algorithm registration systems differ across these units of analysis.

\subsection{China's Beian: Pre-Market Approval and Supervisory Gatekeeping}
The Chinese algorithm registration system focuses on the private sector, emphasizing ex ante, pre-market regulatory control. Compliance is mandatory and stringent: failure to register or meet the requirements can prevent deployment.

\subsubsection{Definition and Key Aims} 

China's algorithm regulatory framework, known as the 算法备案系统 (\textit{Beian system}), is a mandatory compliance mechanism requiring providers of algorithmic services to place key information about their systems on record with regulators. The system encompasses both filing (备案) and registration (登记);\footnote{In China's algorithm governance framework, filing (备案) and registration (登记) are related and partially overlapping compliance obligations. Registration follows a comparatively streamlined process and primarily serves record-keeping and supervisory purposes without prior substantive approval. Filing entails a more complex procedure, including submission of a security self-assessment report and public disclosure of algorithmic mechanisms, and results in the issuance of an official Beian number (备案号).} given their procedural overlap, this paper refers to both collectively as \textit{Beian}. The Beian system operates under a top-down supervisory logic~\cite{Zhang2024PromisePerils, sheehan2024tracing,ComparingApplestoOranges}: algorithms may only operate once regulators obtain structured oversight through Beian.


The concept and practice of Beian predates algorithmic governance and has long been embedded in Chinese administrative practice~\cite{zhang2023,ChinaMediaProject2022}. Its extension to concern algorithmic governance began with the 2021 \textit{Guiding Opinions on Strengthening the Comprehensive Governance of Internet Information Service Algorithms} (hereafter Guiding Opinions). The register did not emerge fully formed, but has taken shape through a gradual broadening of regulatory focus; first zeroing in on algorithmic recommendation systems that curate, rank, or push information to users, and later branching out to cover large-scale generative models focusing on their technological capacities and the risks arising from them.


This evolution also explains \textit{the coexistence of two distinct yet interlocking tracks.} Algorithmic service Beian is keyed to concrete service functions and looks outward to behavioural effects and social risks, including disruptions to social stability arising at the point of interaction with users. Large model Beian, by contrast, looks inward to the technical substrate itself, capturing the scope, scale, and controllability of foundation models (e.g., training data and associated safety measures) before or as they are made available to the public or to third-party developers. The former focuses on service-level algorithmic behaviour and its effects on users, while the latter focuses on latent technical capability and structural risk of the models themselves. This distinction is examined further in \S\ref{Informationrequired}.

Since the initial proposal to establish an algorithmic Beian system in 2021, it has been clearly defined as part of China's broader AI regulatory framework. Registration under the Beian system serves as a mandatory prerequisite for market access, aiming to ensure that algorithmic service providers comply with national regulations before deploying their systems. According to an official government document, regulators also use Beian to identify and address illegal or non-compliant practices.\footnote{“对..备案等工作中发现的..涉算法违法违规行为，予以严厉打击，坚决维护互联网信息服务算法安全” — translation: “Severely punish illegal or non-compliant algorithm-related activities discovered through Beian.” Taken from Guiding Opinions on Strengthening Comprehensive Governance of Algorithms in Internet Information Services} Beyond supervisory gatekeeping control, Beian also functions as an industrial policy instrument. Several local governments link subsidies to successful registrations, using Beian to encourage compliance and attract AI firms. For example, Shenzhen offers subsidies of up to \textit{RMB 1 million per registered algorithm}~\cite{sz_gov_ai_algorithm_support_2024}, indicating that Beian simultaneously serves regulatory and developmental objectives in practice.

\subsubsection{Legal Basis}
\label{sec:chinalegal}

\begin{figure*}[h]
    \centering
    \includegraphics[width=0.8\linewidth]{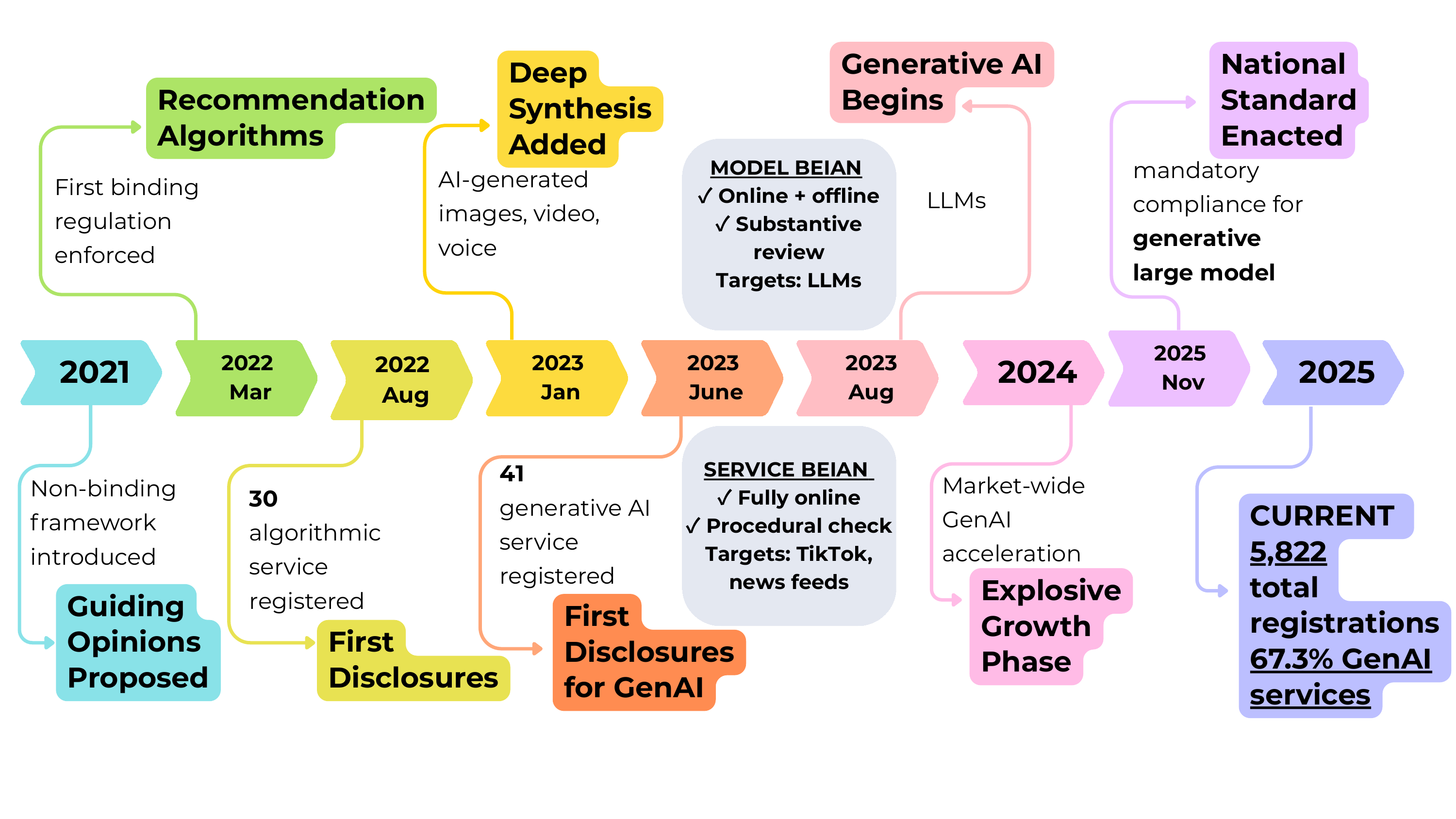}
    \caption{Evolution of China's algorithm registration regime (2021–2025). This period illustrates a shift from non-binding guidance to binding algorithmic regulation, alongside the integration of generative AI services into the registration framework and a broader move toward mandatory national standards. These regulatory developments have been accompanied by significant growth in market participation: as of November 2025, there were 5,822 total registered algorithms, of which 67.3\% relate to generative AI services.}
    \label{fig:timeline}
\end{figure*}

The legal foundations of the Beian system lie in an expanding set of application-specific administrative regulations, reflecting China's agile and adaptive approach to digital governance~\cite{Agileanditerative, AdaptiveSovereignty}. The system developed incrementally: high-level policy guidance in 2021, binding departmental rules issues by the Cyberspace Administration of China (CAC) in 2022–2023, and detailed technical standards effective from 2025 (See Figure \ref{fig:timeline}). This gradual development allowed the Beian system to expand in scope, helping to explain the emergence of a de facto dual registration system for algorithmic services and large generative models.

The first binding obligation appeared in the 2022 \textit{Provisions on the Management of Algorithmic Recommendations in Internet Information Services} (hereafter Algorithmic Recommendations Provisions). It constituted a set of sector-specific administrative rules within China's legal hierarchy. Formally classified as \textit{guīdìng} (规定), they are normative documents issued by State Council ministries or commissions---here, the CAC---pursuant to delegated regulatory authority. Although positioned at a relatively low hierarchical level compared with national laws or administrative regulations enacted by the State Council, such provisions are legally binding within their designated regulatory domain and serve as the primary operational instruments through which administrative agencies structure compliance obligations and enforcement in practice. These rules require providers of algorithmic services to register information about their algorithms. To operationalize compliance, the CAC established a centralized online registration portal (\href{https://Beian.cac.gov.cn/}{https://Beian.cac.gov.cn}) and began publicly disclosing approved registrations in August 2022.

Subsequent regulations expanded the scope of the Beian mechanism beyond recommendation algorithms to encompass emerging AI technologies. \textit{The Provisions on the Administration of Deep Synthesis of Internet-based Information Services} (hereafter Deep Synthesis Provisions) extended the Beian obligations to AI-generated images, video, and voice content, while the\textit{Interim Measures for the Administration of Generative Artificial Intelligence Services} (hereafter Generative AI Measures) brought large language models and other generative systems squarely within the Beian framework. Table~\ref{legalcomparison} in the Appendix provides a comparison of these regulatory provisions. \textit{A de facto dual registration system emerged}, distinguishing algorithmic service Beian from large model Beian. Technical standardization has reinforced the latter track: the 2025 national standard \textit{Basic Safety Requirements for Generative Artificial Intelligence Services} (hereafter Safety Requirements) formalized security obligations and now functions as a de facto compliance benchmark for large-model registration. 

Regulatory authorities lead the verification and audit processes, which constitute mandatory prerequisites for market access. Failure to complete Beian can result in serious consequences for enterprises, ranging from being unable to launch services and forced takedowns to fines and, in extreme cases, criminal liability for responsible individuals. Some app stores, following regulatory directives, have already removed non-compliant apps and now require an algorithm registration number for new listings~\cite{sohu_aigc_violation_cases_2024,rmrb_ai_applications_2025}.

\subsubsection{Scope and Coverage}
\label{sec:chinascope}
The Beian system applies to private enterprises and companies providing algorithmic and information services. Beian is mandatory for algorithmic service providers whose systems possess ``public opinion properties'' or the ``capacity for social mobilization''. Although the regulations do not explicitly define these criteria~\cite{lathamwatkins_chinas_new_ai_regulations_2023}, in practice, Beian applies to all public-facing AI services in China, including chatbots, image generators, and user-accessible APIs, regardless of whether they are developed in foreign jurisdictions~\cite{dtalliance_ai_regulation_eu_china_2024}.

\begin{figure*}[h]
    \centering
    \begin{subfigure}[b]{0.49\linewidth}
        \centering
        \includegraphics[width=\linewidth]{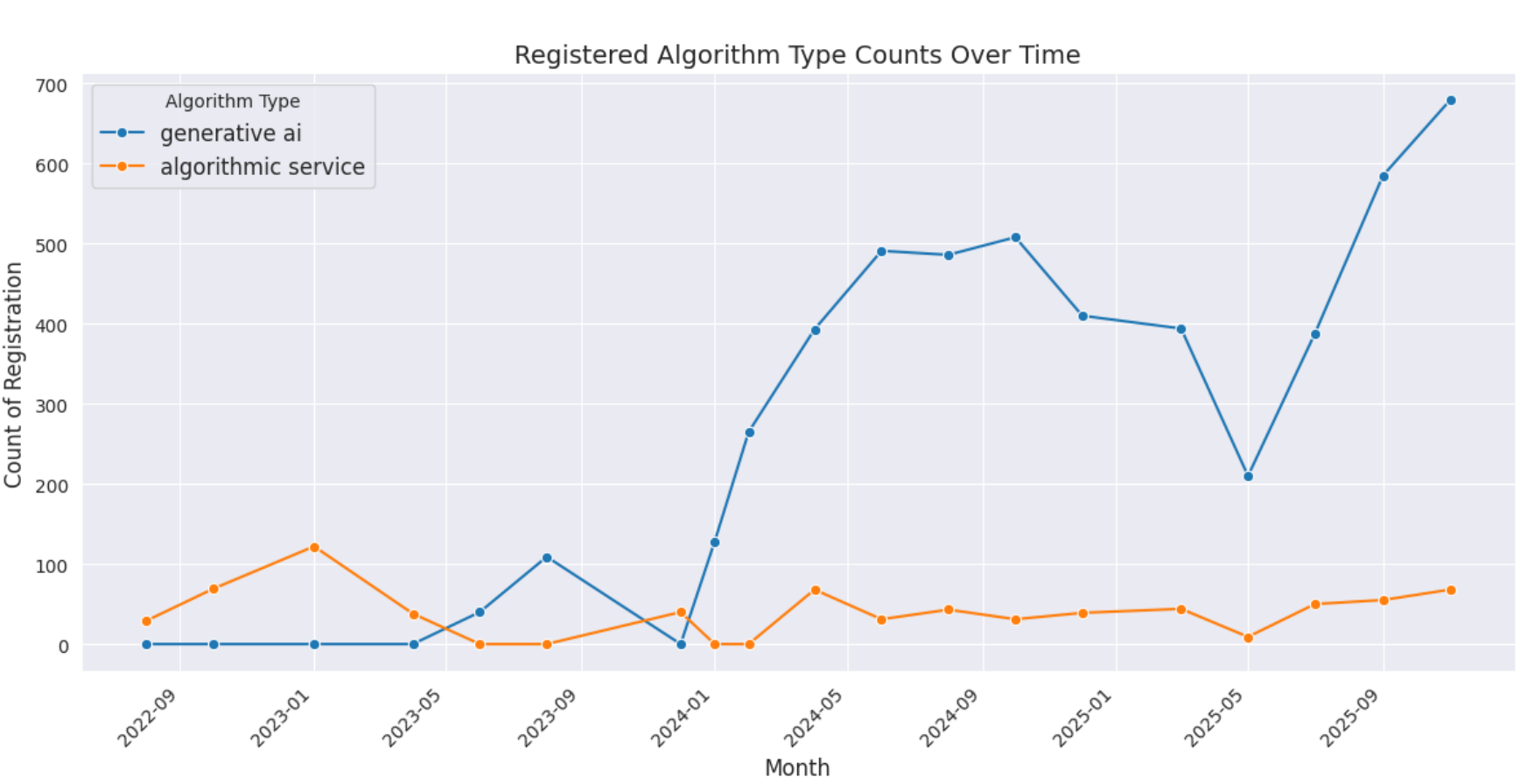}
        \caption{Publicly disclosed registrations of algorithmic recommendation and generative AI services by disclosure period in China.}
        \label{fig:registered_algorithm_type}
    \end{subfigure}
    \hfill
    \begin{subfigure}[b]{0.49\linewidth}
        \centering
        \includegraphics[width=\linewidth]{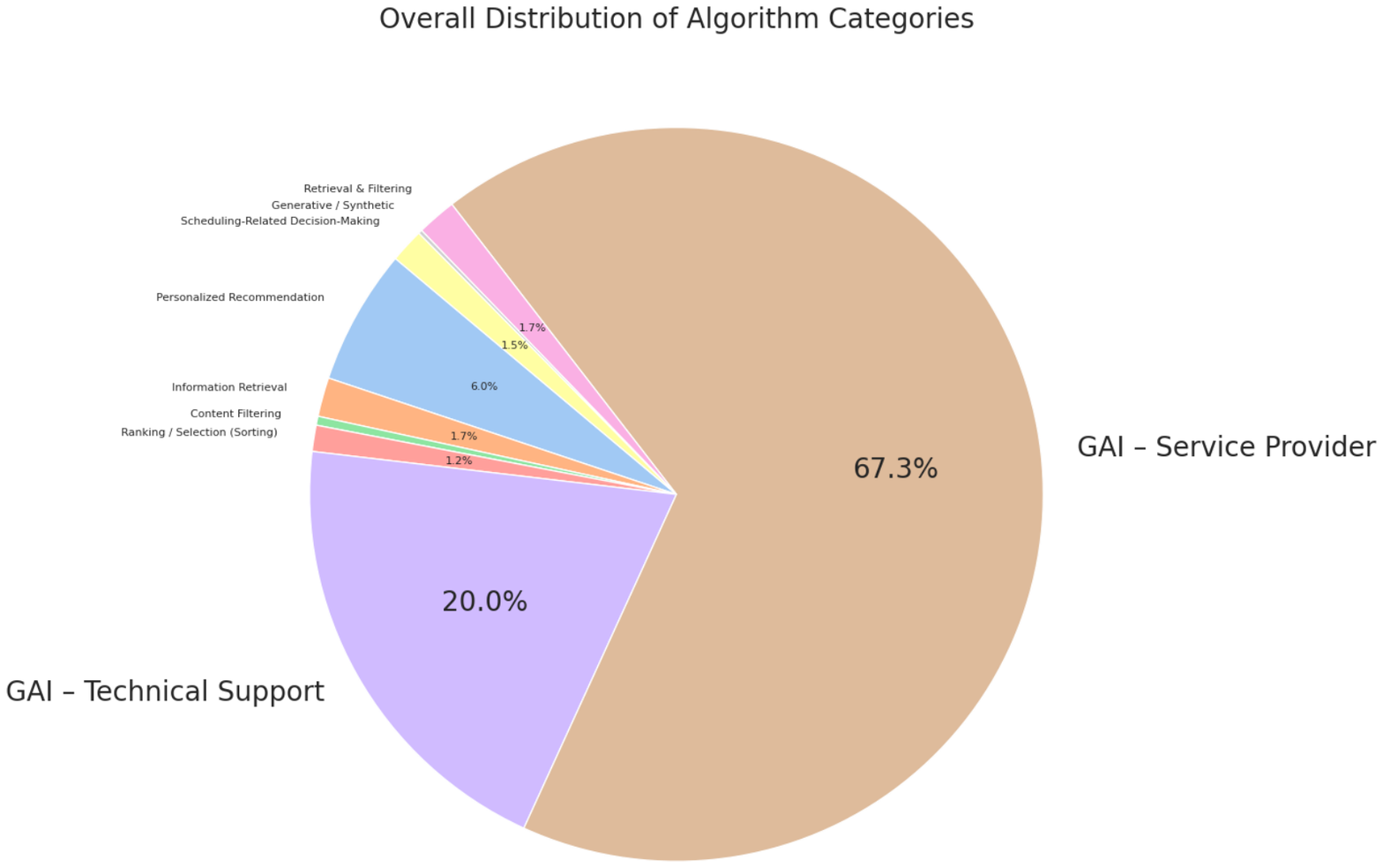}
        \caption{Overall distribution of algorithm categories. Generative AI – Service Provider accounts for 67.3\%, Generative AI – Tech Support for 20.0\%.}
        \label{fig:pie_chart}
    \end{subfigure}
    \caption{Algorithm registration trends and category distribution in China.}
    \label{fig:combined_figures}
\end{figure*}

Since its introduction, the Beian system has expanded in phases. Initially focused on algorithmic services, it later incorporated generative AI, creating a \textit{dual-track registration framework}: one for algorithmic services emphasizing behavioral and social effects, and one for large models emphasizing technical capabilities. Analysis of registration data reveals clear phase shifts aligned with these milestones. Early disclosures of algorithmic recommendation services in 2022 were limited. From mid-2023 onward, registrations surged, driven almost entirely by generative AI (Figure \ref{fig:registered_algorithm_type}). By 2025, generative AI accounted for nearly 88\% of all registrations, 67.3\% for generative AI services and a further 20\% for generative AI technical support, reflecting its rapid commercialization (Figure \ref{fig:pie_chart})

As of November 2025, the registry includes 5,822 unique entries from 3,787 entities. Analyzing these registered entities provides insight into China's AI ecosystem. Most entities (76\%) have registered only a single algorithm, indicating that the majority of participants are small-scale actors. In contrast, repeated registrations are concentrated among a few established technology leaders. The top five registered entities are: Tencent (33), ByteDance (23), Baidu (19), SenseTime (16), and Xiaobing (originally Microsoft Research Asia, 14). DeepSeek, despite launching a high-profile large language model on January 2025, has only two registrations, with its first appearing in the June 2024 disclosure batch, seven months before its public launch. The registration data reveal a long-tail distribution: a small number of established firms lead the industry, while a large number of smaller actors contribute to its diversity and rapid growth. These patterns, made visible through the Beian system, provide a useful window for both regulators and external observers into the structure, scale, and competitive dynamics of China's AI ecosystem.

\subsubsection{Information Requirements}
\label{Informationrequired}

All registrations are submitted through the Online Beian System and require three core categories of information: \textit{algorithm entity information, algorithm information, and product and function information}. Algorithmic service Beian follows a standardized, fully online process. By contrast, large-model Beian combines online submission with offline engagement with local CAC offices, involving iterative consultations and testing despite the absence of an explicit legal basis for this hybrid process. We summarize the key differences between algorithmic service Beian and large-model Beian in Table~\ref{tab:Beian_comparison} in the Appendix.

\textit{Algorithm Entity Information} requires submission of a business license, the legal representative's identification, the algorithm responsible person's identification and proof of employment, a letter of commitment, and the ICP record or license, which certifies that websites or apps operating in mainland China comply with local regulatory requirements. \textit{Algorithm information} includes both basic descriptors and detailed technical attributes. While extensive technical data is submitted, public disclosure is limited to high-level summaries such as algorithm category, entity name, application product, purpose, and Beian number. A central requirement is the algorithm security self-assessment report, completed using CAC templates. Importantly, rather than providing a one-size-fits-all template, the system offers category-specific self-assessment templates tailored to the unique features and risks of different types of algorithms. The algorithm categories for which detailed templates are provided under Algorithmic service Beian include:

\begin{enumerate}
\item \textit{Personalized Recommendation} – Algorithms that provide individualized content or product suggestions.
\item \textit{Information Retrieval} – Algorithms that collect, index, and return relevant information from large datasets.
\item \textit{Content Filtering} – Algorithms that detect and filter harmful, illegal, or inappropriate content.
\item \textit{Ranking/Selection(Sorting)} – Algorithms that determine the order or priority of items presented to users.
\item \textit{Scheduling-Related Decision-Making} – Algorithms that optimize allocation, delivery, or operational decisions.
\item \textit{Generative AI – Technical Support} – Algorithms that provide generative AI capabilities as technical support for enterprise clients.
\item \textit{Generative AI – Service Provider} – Algorithms that provide generative AI services directly to end-users.
\end{enumerate}

All algorithm types are required to implement risk mitigation measures, but the focus differs based on algorithm function and impact: recommendation and ranking systems focus on bias, manipulation, and user autonomy, while generative AI
emphasizes national security, misinformation, and harmful outputs. Table \ref{tab:self-assessment difference} in the Appendix summarizes the different requirements across algorithm categories, including: Risk Mitigation, User Rights Protection, Content Ecosystem Governance, Decision Evaluation, Content Diversity Safeguards, Model Security, and Data Security.

\textit{Product and Function Information} links algorithms to their deployment contexts. Registrants must specify product details, access paths to algorithmic functions, descriptions of functionality along each path, platform information, monthly active users, personalization controls, and youth protection measures.

For large-model Beian, additional materials are required, including model service agreements, data annotation rules, keyword blocking lists, and evaluation test sets. The 2025 Safety Requirements mandates lifecycle-wide security controls, including training-data audits, output monitoring, and post-deployment risk management. This structured, category-specific requirement demonstrates a sophisticated understanding of both the technical characteristics of different algorithms and their social, economic, and regulatory implications. At the same time, the practical ability of authorities to fully verify such detailed submissions remains uncertain, suggesting a potential tension between regulatory ambition and implementation capacity.

\subsubsection{Disclosure and Maintenance} 
\label{sec:chinadisclosure}

Both algorithmic service providers and large model providers are required to keep their registration information up to date. Any changes to previously filed information must be reported through the prescribed change procedures within \textit{ten working days}. Where an algorithm recommendation service is terminated, providers must complete the relevant cancellation procedures within \textit{twenty working days} of service cessation.

While the registration regime requires the submission of detailed technical information about the algorithm, the relevant product, and evaluation results, it takes a very limited approach to public disclosure. The full technical details submitted during registration are not made publicly available. Instead, enterprises are required to provide a Public Disclosure Content (公示内容) section, which includes specified information written in clear, non-technical language. Public access to registration information is further limited by the method of disclosure.\footnote{Specified information includes algorithm name, algorithm category, entity name, application name, main purpose, and Beian number.} There is no dedicated, web-accessible interface that allows users to search, filter, or systematically explore registered algorithms. Instead, the CAC periodically releases batches of approved registrations as downloadable Word documents on its websites. Users must download and manually compile these files to analyze or query the data. Taken together, these design choices suggest that the Beian system prioritizes regulatory oversight and administrative control over broad public transparency.

\subsection{UK's ATRS: Self-Declared Transparency Without Review and Verification}
In contrast, the UK system focuses primarily on the public sector and aims to promote transparency and public accountability. Registration is self-declared, with no formal review or regulatory enforcement to influence deployment, reflecting a framework that prioritizes openness and information-sharing over mandatory oversight.


\subsubsection{Definition and Key Aims}

The UK's Algorithm Register is an online repository\footnote{The term `repository' is used here to indicate a structured collection of registration information. It does not imply that the underlying data, model, weights, or source code are deposited or otherwise made available through the register.} through which public sector organizations record and publish information about how they use algorithmic tools. It operates under the Algorithmic Transparency Recording Standard (ATRS), which provides a common template for documenting algorithmic systems used across government. Unlike China's Beian System, the ATRS does not function as an approval mechanism or pre-deployment gate. Instead, it is a post-hoc transparency and accountability framework focused on disclosure rather than authorization. Originally introduced as a voluntary standard between 2021 and 2023, ATRS became mandatory for central government departments and certain public bodies (such as arm's-length bodies\footnote{Arm's-length bodies are UK public organizations that operate independently from direct government department control but deliver public services, functioning with operational autonomy while remaining accountable to a sponsoring minister and Parliament.}) from February 2024~\cite{UKproinnovation,govuk_blueprint_modern_digital_gov_2025}.

Despite its mandatory status, ATRS is not backed by formal enforcement mechanisms. \changeyp{This appears partly due to the scope of ATRS: as a standard only targeting public sector bodies, it operates through internal governance norms rather than external enforcement mechanisms.} Failure to register an algorithmic tool does not prevent deployment of an algorithmic system, nor does it trigger any formal consequences. In practice, compliance relies on internal governance processes, adherence to organizations' government standards and best practice. The ATRS forms part of the UK government's broader digital reform agenda~\cite{govuk_blueprint_modern_digital_gov_2025}, supporting the responsible use of AI in public services. It aims to enable the government to expand AI capability, deploy AI ethically, and ensure transparency in algorithmic decision-making. By standardizing documentation, the register promotes public trust, accountability, and administrative efficiency~\cite{govuk_algorithmic_transparency_2025}. \changeyp{In addition, it can also support public procurement as a lever for shaping public sector AI in line with public interests~\cite{WildWestProcurement,GatekeepingPublicProcurement}, by providing further clarity to third-party suppliers regarding public sector transparency requirements.} According to publicly available ATRS guidance~\cite{govuk_algorithmic_transparency_2025}, its key objectives include:




\begin{itemize} 
    \item \textit{Public Trust:} Drive public understanding and trust in algorithmic tools;
    \item \textit{Accountability:} Enable senior responsible owners to take meaningful accountability;
    \item \textit{Learning \& Sharing:} Share good practice and innovative use cases, and learn from peers;
    \item \textit{Administrative Efficiency:} Reduce administrative burden through proactive information publication;
    \item \textit{Public Procurement:} Provide clarity to third-party suppliers on transparency requirements.
\end{itemize} 

\subsubsection{Legal Basis}
\label{sec:uklegal}
The ATRS is implemented through administrative guidance and standards rather than a statutory duty, emerging from the UK government's broader digital reform agenda~\cite{govuk_blueprint_modern_digital_gov_2025}, which emphasizes harnessing AI responsibly for public benefit while ensuring transparency and public trust. It constitutes a non-statutory administrative requirement: public bodies are expected to comply, but there is no formal legislation or regulation that imposes legal consequences for non-compliance. Although \changeyp{no dedicated legal instrument underpins ATRS}, its design and implementation draw on existing legal and policy frameworks to guide practice and ensure consistency with broader transparency and governance expectations. These frameworks provide context and operational guidance rather than forming the legal source of the requirement itself:


\begin{enumerate}
\item \textit{Public sector transparency obligations and commitments}: These broader policy norms and institutional expectations drive the creation of ATRS itself. They provide the rationale for requiring public bodies to document and disclose information about their algorithmic tools.
\item \textit{Freedom of Information Act 2000 (FOIA)}: FOIA provides a reactive framework for public access to government information. ATRS adopts the logic of FOIA exemptions to decide what information is too sensitive to publish proactively. In other words, FOIA informs what should not be disclosed.
\item \textit{Data protection and privacy law}: Similarly, data protection rules determine which personal or sensitive data must be withheld from public disclosure.
\end{enumerate}


ATRS records are maintained under the responsibility of a senior responsible owner within each public sector organization, who is accountable for ensuring that records are accurate, complete, and kept up to date. While this role establishes internal accountability, approaches to evaluating and reporting algorithmic tools may vary across organizations. Because full ATRS records are published online, they also enable external scrutiny: members of the public, civil society organizations, and investigative journalists can review the information, raise concerns, and request clarification. For example, ~\cite{amnesty_automated_racism_2025} illustrates how civil society used registration data to investigate and critically assess government use of algorithmic systems. 



\subsubsection{Scope and Coverage} 
\label{sec:ukscope}

\begin{figure}[t]
    \centering
    \includegraphics[width=0.9\linewidth]{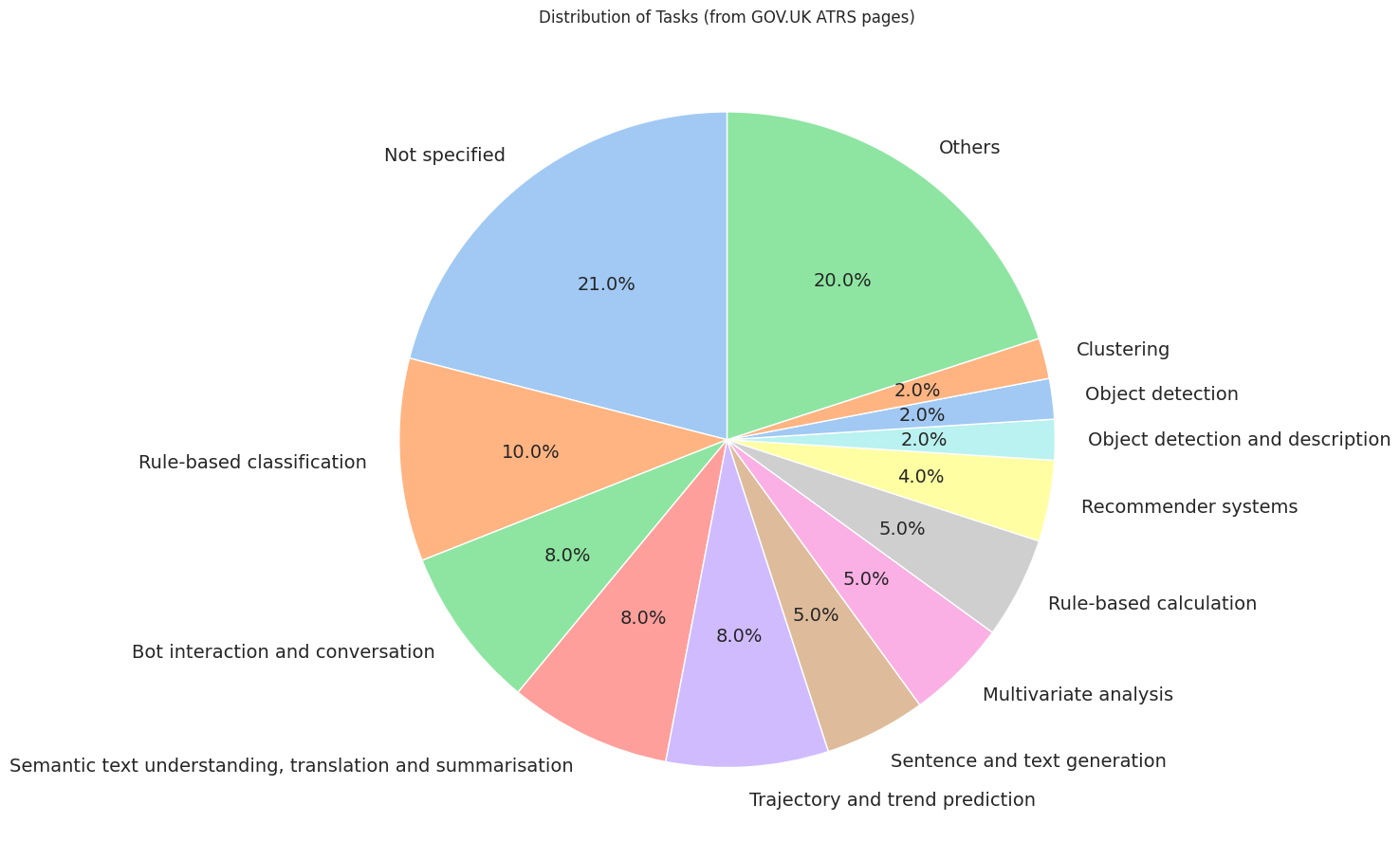}
    \caption{Algorithm types and tasks under ATRS.}
    \label{fig:atrs_by_types}
\end{figure}

The ATRS applies to central government departments and certain public bodies using algorithmic tools that either significantly influence decision-making with public impact or directly interact with the general public. Tools used exclusively for internal analysis, policy-making, or administrative purposes that do not materially affect citizens are generally out of scope, though organizations are encouraged to register in cases of uncertainty to promote transparency and best practice. For example, the Department for Science, Innovation and Technology (DSIT) has registered its internal Chatbot 24/7, which provides guidance on internal HR queries~\cite{ukgov_dsit_ask_ops_chatbot_2024}. The ATRS defines an ``algorithmic tool'' as a broad category encompassing a wide range of AI applications, from simple statistical models to complex algorithms. Approximately 10\% of registered tools are rule-based classification systems (See Figure \ref{fig:atrs_by_types}).

    

\begin{figure}[t]
    \centering
    \includegraphics[width=0.5\linewidth]{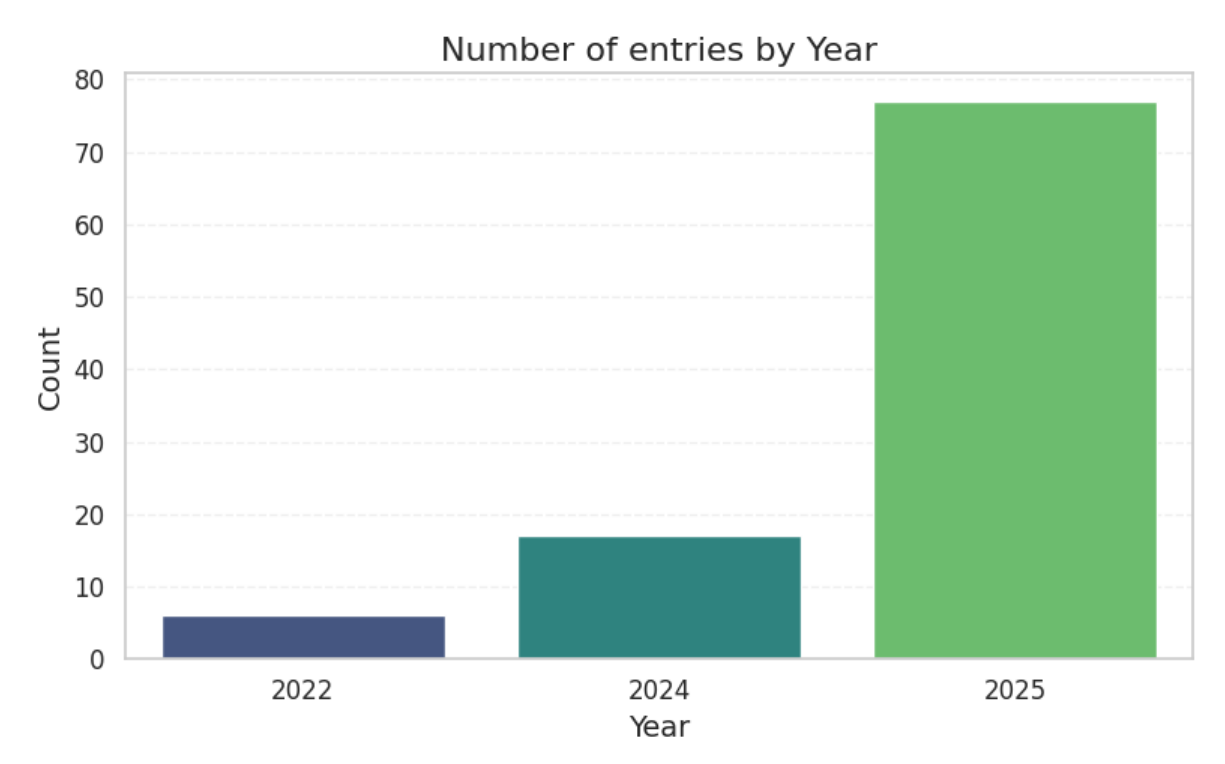}
    \caption{Registrations under ATRS by year (2022–2025).}
    \label{fig:atrs_by_year}
\end{figure}

The ATRS was first launched in November 2021 and piloted across a variety of public sector organizations~\cite{piloting_algorithmic_transparency_2022}. Early pilots focused on influential operational algorithms, such as the QCovid risk-stratification model developed by the Department of Health and Social Care and NHS Digital~\cite{covidrisk}. Prior to mandatory implementation, adoption was limited, with fewer than ten tools registered in 2022; this may reflect both the early stage of the ATRS and the relatively smaller number of algorithmic systems in use at the time. Since February 2024, the ATRS has been mandatory for central government bodies. Following the mandate, the number of published records increased substantially, reaching roughly eighty by 2025 (See Figure \ref{fig:atrs_by_year}). Departments with the largest number of registered tools include the Department for Education, the Department for Science, Innovation and Technology, the Department of Health and Social Care, the Department for Business and Trade, and the Ministry of Justice. Several local police forces have also recorded risk-assessment tools (See Figure \ref{fig:uk_register_by_department} in the Appendix).

\subsubsection{Information Requirements}

Since its launch in 2021 as a voluntary registration mechanism, the ATRS has evolved and been shaped by lessons from its pilot phase~\cite{rtau_engaging_public_algorithmic_transparency_2021} and external engagement, including public focus groups and surveys examining attitudes toward algorithmic transparency in the public sector~\cite{britainthinks_transparency_simplicity_2021}. This feedback informed the development of a tiered reporting approach designed to present information clearly and simply while maintaining meaningful transparency~\cite{britainthinks_transparency_simplicity_2021}. Tier 1 ``provides minimal, high-level information about an algorithmic tool, suitable for public disclosure. It offers a general overview without revealing sensitive technical details" while Tier 2 ``contains more comprehensive information, explaining how the tool works, its purpose, and its impact." To promote consistency in reporting, organisations are provided with guidance and a standardised template. Unlike China's algorithm-specific registration template, the ATRS provides a generic template applicable to all algorithmic tools. Importantly, organisations registering an algorithmic tool are required to complete both tiers. During the ATRS pilot phase, however, only Tier 1 information was made publicly available, while Tier 2 disclosures were intended to be shared on request~\cite{britainthinks_transparency_simplicity_2021}. In the current implementation, however, both Tier 1 and Tier 2 records are publicly disclosed by default. The rationale for this move toward full public disclosure of both tiers is unclear. Table~\ref{tab:ATRS_tiers_two_columns} in the Appendix presents a detailed breakdown of the information required in each tier.



\subsubsection{Disclosure and Maintenance}
\label{sec:ukdisclosure}


ATRS must be regularly reviewed and updated to reflect any substantive changes to algorithmic tools. Substantive changes may include~\cite{govuk_algorithmic_transparency_2025}:
\begin{enumerate} 
    \item A pilot tool moving into full production,
    \item The introduction of new datasets for training, validation, or refinement,
    \item Modifications to the broader operational process in which the tool is deployed.
\end{enumerate} 
These updates should be made promptly and reflected in the publicly accessible online repository. However, it is ultimately left to the organization to determine what constitutes a substantive change. Unlike China's approach, there is no legally specified timeframe for reporting such changes, nor are there explicit consequences for non-compliance. The web-accessible repository includes robust search functionality, allowing users to locate algorithms by organization, organization type, function, technical capability, deployment phase, region, and publication date. Users can also subscribe to email notifications for updates and submit feedback via an optional form, providing both quantitative ratings and qualitative suggestions (See Figure \ref{fig:web_interface} in the Appendix). Compared with China's minimal, static disclosure model, the UK's approach makes information more discoverable. Yet, in practice, meaningful transparency depends on whether citizens, media, and civil society can actively use this information and how useful it is for their scrutiny~\cite{DisclosurebyDesign,Miricrim,NullCompliance}.

An illustrative example of how ATRS reporting and maintenance requirements play out in practice is the registration for the Hampshire and Thames Valley Police Domestic Abuse Risk Assessment Tool (DARAT), published in 2022 under the pre‑deployment category~\cite{ukgov_hampshire_thames_valley_police_darat}. DARAT is a high-stake system designed to help `police officers effectively grade the risk of future harmful incidents of domestic abuse'~\cite{ukgov_hampshire_thames_valley_police_darat}. The Tier 1 and Tier 2 ATRS records for DARAT provide the purpose and scope of the tool, the detailed data sources used in its construction, and an extensive list of 46 identified risk factors. Despite this level of detail, our analysis of the DARAT record also highlights key limitations in how the ATRS functions as a transparency mechanism. The record does not include an impact assessment; instead, it notes that this would be provided prior to deployment. However, as of 2025, the DARAT entry has not been updated to include the promised impact assessments or reflect changes in its design and deployment plans. Only through information from other sources is it known that DARAT has been put on hold~\cite{amnesty_automated_racism_2025}.

\section{The Design and Implementation of the Algorithm Registers Shape its Role in AI Governance}

Taken together, the preceding analyses show how variation in the design and implementation of algorithm registers shapes their governance-related effects. In China, Beian is a pre-market approval and regulatory verification system that is still evolving with technological progress and provides limited public disclosure. In the UK, by contrast, registration of central government algorithmic tools is disclosure-oriented, self-reported, and unverified. While information is publicly accessible, oversight relies primarily on external actors rather than direct state enforcement. The more detailed discussion of China's Beian is reflective of the broader range of governance aspects it entails.


These variations demonstrate that there is no single uniform model of an ``algorithm register.'' Rather, the design choices concerning scope, disclosure, and verification requirements shape what regulatory functions a register might perform. Our comparative analysis therefore moves beyond debates about whether registers ``deliver transparency'' (see \S\ref{sec:background}) based solely on the information they record. Instead, we \textbf{emphasize the importance of examining the processes enabled by design choices and the institutional contexts in which registers are embedded}, grounding our understanding of algorithm registers in their concrete design and implementation. 

\changeyp{In what follows, we identify three core roles that algorithm registers can play and show how each is enabled or constrained by the specific analytical dimensions examined in §\ref{sec:compare}: (i) how scope and disclosure dimensions shape a  register's capacity for \textit{ecosystem-level understanding}; (ii) how verification and enforcement mechanisms determine the extent of \textit{regulatory oversight}; and (iii) how the evolution of scope and information requirements reveals a register's potential as \textit{expandable regulatory infrastructure}.}

 \paragraph{Registers Enable Ecosystem-Level Understanding Even with Poor Individual Records}


Algorithm registers aggregate individual registrations into a collective information infrastructure that enables ecosystem-level understanding of algorithmic development and deployment. \changeyp{As shown across the disclosure dimensions (\S\ref{sec:chinadisclosure} and \S\ref{sec:ukdisclosure})}, individual disclosures are often partial, strategically vague, or minimal; \changeyp{yet their value lies not just in what any single entry reveals about a particular algorithmic system, but in enabling observation of aggregate patterns across actors, sectors, and regions.}

China's Beian system illustrates how this function can operate even under conditions of limited disclosure and restricted accessibility. Although each registration provides little more than basic functional and organizational information and public access is technically fragmented, the aggregation of registration data nonetheless reveals important structural features of China's AI ecosystem. These features include the scale and growth of generative AI deployment (see Figures \ref{fig:registered_algorithm_type}, \ref{fig:line_chart}, \ref{fig:pie_chart}), the distribution of activity across firms (see \S\ref{sec:chinascope}), and the geographic concentration of development in major innovation hubs~\cite{triviumchina_seeking_next_deepseek_2025}. Internally, registers help authorities monitor trends, identify emerging risks, and guide policy; externally, they allow researchers and analysts to study market dynamics. In this sense, registers can serve as instruments of system-level understanding even when they are limited as tools for auditing or evaluating individual systems.

Similarly, the UK's ATRS provides insight into the types of algorithmic tools used in the public sector and how they are deployed \changeyp{(See \S\ref{sec:ukscope})}. However, currently there is no register in Europe for private-sector AI systems. The \changejs{EU} AI Act has registration requirements for high-risk AI systems\footnote{The EU AI Act mandates an EU database for high-risk AI systems (Annex III) under Article 71} which, depending on implementation specifics, may support broader understandings of AI deployment across the economy
~\cite{woersdoerfer_eu_ai_act_2025,Svitych21042025}.

\paragraph{Oversight Depends on Mandatory Registration, Verification, and Consequences}

A central difference between China's and the UK's algorithm registers, rooted in the verification and enforcement mechanisms examined in legal basis dimensions \changeyp{(See \S\ref{sec:chinalegal} and \S\ref{sec:uklegal}),} is the degree to which they enable enforceable oversight and control, that is, the ability of regulators to verify registrations, intervene and impose consequences for non-registration or failed review. In China, registration under the Beian system is a prerequisite for market access. Detailed technical submissions, coupled with mandatory updates, allow regulators to assess whether providers meet substantive AI governance requirements. Deployment without a Beian number is illegal, embedding oversight directly into the registration process.

By contrast, the UK's ATRS operates within a self-reporting framework. Although registration is formally mandatory for central government bodies, there is no external verification or formal consequences for non-compliance. Because ATRS currently covers only government organizations, the lack of consequences for non-compliance should be interpreted differently for registers targeting private-sector actors. Nevertheless, a UK House of Lords report noted that, to ensure the register remains reliable, accurate, and up to date, regular review of entries and some form of penalty for incomplete or missing submissions would be necessary~\cite{hl_tech_rules_2022}.



\paragraph{Registers as Expandable Regulatory Infrastructure}

China's Beian system, as traced through the legal basis and scope dimensions in §\ref{sec:chinalegal} and §\ref{sec:chinascope}, exemplifies a form of ``regulatory scaffolding''~\cite{sheehan2023_chinese_ai_gov}: a reusable infrastructure that allows regulators to layer successive requirements and interventions while incrementally building bureaucratic know-how and regulatory capacity. Initially introduced to cover algorithmic recommendation services, the Beian system was later extended to generative AI systems, illustrating how a single registration mechanism can serve as a foundation for expanding governance measures. Specialized self-assessment templates for different algorithm types (See Table \ref{tab:self-assessment difference}) and safety requirements for large models show that the Beian system accommodates additional obligations without creating entirely new regulatory mechanisms. Similarly, requirements for labelling AI-generated content were integrated into existing registration procedures in 2025~\cite{geopolitechs2025}. As of the time of writing this paper (Dec 2025), China has published its Interim Measures for the Management of Anthropomorphic AI Interactive Services (人工智能拟人化互动服务管理暂行办法, draft for public comment), which again stress the requirement of registration~\cite{cac_2025_ai_hrenhua}. By gradually expanding its scope and adding specific requirements for new algorithm types, Beian demonstrates how a registry can function as an expandable regulatory backbone that evolves alongside technological development.

By contrast, the UK's ATRS currently plays a more limited infrastructural role. Although it has transitioned from a pilot initiative to a mandatory scheme, its scope, focus, and information requirements have remained largely unchanged. To build on the foundation of ATRS and increase its adaptability, regulators could periodically assess emerging technological risks to determine whether ATRS should be expanded to cover additional sectors, actors, or include specific requirements for particular algorithm types. Moreover, proposals to link the registry with incident-reporting mechanisms would further harness its infrastructural role, enabling regulators to trace observed harms back to specific systems or models and implement more targeted interventions~\cite{mckernon2024aimodelregistriesfoundational}.

\section{Limitations and Future Work}
The comparative scope of this analysis was limited to China and the UK. Algorithm registers in other jurisdictions may exhibit additional forms of variation, and we do not claim nor seek to capture all possible register configurations. The Chinese and UK cases differ substantially in their institutional arrangements and regulatory objectives, enabling us to surface a wide range of design choices and governance functions relevant to support the paper's aim of advancing and encouraging a more holistic understanding of algorithmic registers. Future work could extend this analysis to additional jurisdictions in order to further explore and refine these observations, and indeed, the paper sets up a comparative framework for doing so.

Moreover, drawing primarily on publicly available materials, the analysis focuses on how algorithm registers in China and the UK are formally designed and intended to function, rather than how they operate in day-to-day practice. In the case of China, for example, we cannot easily observe how regulators actually verify technical submissions, whether they have sufficient capacity to assess the information provided, or what verification procedures are followed in practice. Nevertheless, examining registers at the level of design (as we did) remains valuable, as it clarifies key variations and the governance functions these systems are intended to perform. Future research could broaden the empirical basis of this work through methods such as interviews or field studies, in order to better understand how design choices translate into operational practice and with what consequences.

\section{Conclusion}
This paper advances a more holistic understanding of algorithmic registers. 
Through analysis of China and the UK, we show that algorithm registration systems can differ substantially, resulting in distinct governance functions such as transparency, compliance verification, and market access approval. 
Rather than focusing narrowly on transparency, the analysis highlights the importance of considering how registers are designed, implemented, and embedded within broader regulatory systems. 
In particular, we highlight that factors such as scope, disclosure requirements, verification processes, and enforcement shape the functions registers perform in practice. 
Understanding registers as configurable governance instruments can clarify how different arrangements give rise to different governance effects across regulatory settings, and support more strategic use of registers in addressing emerging AI risks.

\section{Generative AI Usage Statement}
ChatGPT (based on GPT-5.2) was used only for limited editorial and technical assistance, including citation formatting, table structuring and layout adjustments, identification of relevant documents and literature, code debugging, and grammar and style editing. All outputs  were carefully reviewed and verified by the authors. 

\bibliographystyle{ACM-Reference-Format}
\bibliography{reference}

\appendix
\newpage

\section{Official Government and Regulatory Documents Included in the Analysis}
\label{tab:official-documents}

\begin{table}[H]
\centering
\scriptsize
\caption{Official Government and Regulatory Documents Included in the Analysis}
\begin{tabular}{p{2cm} p{2cm} p{10cm}}
\hline
\textbf{Jurisdiction} & \textbf{Index} & \textbf{Document Title} \\
\hline
UK & UK-1 & Written Questions, Answers and Statements – UK Parliament \\
& UK-2 & Algorithmic Transparency Recording Standard (ATRS\_V4.0\_\_FINAL) Template \\
& UK-3 & Algorithmic Transparency Recording Standard – Guidance for Public Sector Bodies \\
& UK-4 & Algorithmic Transparency Recording Standard (ATRS) Mandatory Scope and Exemptions Policy \\
& UK-5 & Independent report - BritainThinks: Complete Transparency, Complete Simplicity \\
& UK-6 & How Can the Public Sector Be Meaningfully Transparent About Algorithmic Decision-Making? \\
& UK-7 & Engaging with the Public About Algorithmic Transparency in the Public Sector \\
& UK-8 & Algorithmic Transparency Recording Standard: Getting Ready for Adoption at Scale \\
& UK-9 & Algorithmic Transparency Recording Standard – Data Standards Authority \\
\hline
China & CN-1 & 网络安全技术 人工智能生成合成内容标识方法 \\
&  & \emph{Cyber Security Technology — Labeling Method for Content Generated by Artificial Intelligence} \\
& CN-2 & 网络安全技术 生成式人工智能服务 安全基本要求 \\
&  & \emph{Cyber Security Technology — Basic Security Requirements for Generative Artificial Intelligence Services} \\
& CN-3 & 互联网信息服务算法安全自评估报告（排序精选类） \\
&  & \emph{Internet Information Service Algorithm Security Self-Assessment Template (Ranking and Selection Type)} \\
& CN-4 & 互联网信息服务算法安全自评估报告（生成合成类—服务技术支持者） \\
&  & \emph{Internet Information Service Algorithm Security Self-Assessment Template (Generated/Synthetic Type — Technical Service Supporter)} \\
& CN-5 & 互联网信息服务算法安全自评估报告（生成合成类—服务提供者） \\
&  & \emph{Internet Information Service Algorithm Security Self-Assessment Template (Generated/Synthetic Type — Service Provider)} \\
& CN-6 & 互联网信息服务算法安全自评估报告（信息检索类） \\
&  & \emph{Internet Information Service Algorithm Security Self-Assessment Template (Information Retrieval Type)} \\
& CN-7 & 互联网信息服务算法安全自评估报告（个性化推送类） \\
&  & \emph{Internet Information Service Algorithm Security Self-Assessment Template (Personalized Recommendation Type)} \\
& CN-8 & 互联网信息服务算法安全自评估报告（内容过滤类） \\
&  & \emph{Internet Information Service Algorithm Security Self-Assessment Template (Content Filtering Type)} \\
& CN-9 & 互联网信息服务算法安全自评估报告（调度决策类） \\
&  & \emph{Internet Information Service Algorithm Security Self-Assessment Template (Scheduling and Decision-Making Type)} \\
& CN-10 & 《互联网信息服务算法推荐管理规定》答记者问 \\
&  & \emph{Q\&A on the Internet Information Service Algorithmic Recommendation Management Provisions} \\
& CN-11 & 人工智能生成合成内容标识管理办法 \\
&  & \emph{Administrative Measures for the Labeling of Artificial Intelligence-Generated Synthetic Content} \\
& CN-12 & 生成式人工智能服务管理暂行办法 \\
&  & \emph{Interim Measures for the Management of Generative Artificial Intelligence Services} \\
& CN-13 & 互联网信息服务算法推荐管理规定 \\
&  & \emph{Internet Information Service Algorithmic Recommendation Management Provisions} \\
& CN-14 & 关于开展互联网信息服务算法备案工作的通知 \\
&  & \emph{Notice on the Launch of the Internet Information Service Algorithm Filing System} \\
& CN-15 & 互联网信息服务深度合成管理规定 \\
&  & \emph{Provisions on the Administration of Deep Synthesis of Internet-Based Information Services} \\
& CN-16 & 互联网信息服务算法备案系统使用手册 \\
&  & \emph{User Manual for the Internet Information Service Algorithm Filing System} \\
& CN-17 & 关于印发《关于加强互联网信息服务算法综合治理的指导意见》的通知 \\
&  & \emph{Notice on the Issuance of the ‘Guiding Opinions on Strengthening Comprehensive Governance of Algorithms in Internet Information Services’} \\
\hline 
\end{tabular}
\end{table}

\section{Comparison of Algorithm-Related Regulatory Provisions in China}

\begin{table}[H]
\centering
\Small
\begin{tabular}{|p{2cm}|p{4cm}|p{4cm}|p{4cm}|}
\hline
\textbf{Item} & \textbf{Algorithmic Recommendation Provisions} & \textbf{Deep Synthesis Provisions} & \textbf{Generative AI Interim Measures} \\
\hline
\textbf{Promulgating Entities} & 
Cyberspace Administration of China (CAC) — main; Ministry of Industry and Information Technology (MIIT), Ministry of Public Security (MPS), State Administration for Market Regulation (SAMR) & 
CAC, MIIT, MPS & 
CAC — main; National Development and Reform Commission (NDRC), Ministry of Education (MOE), Ministry of Science and Technology (MOST), MIIT, MPS, National Radio and Television Administration \\
\hline
\textbf{Date of Issuance / Implementation} & 
Issued Dec 31, 2021; effective Mar 1, 2022 & 
Issued Dec 11, 2022; effective Jan 10, 2023 & 
Issued Jul 10, 2023; effective Aug 15, 2023 \\
\hline
\textbf{Regulatory Nature / Level} & 
Departmental regulation (formal, stable rule) & 
Departmental regulation (targeting synthetic content) & 
Interim administrative measure (transitional, exploratory regulation) \\
\hline
\textbf{Target Entities for Beian} & 
Algorithmic recommendation service providers with public opinion attributes or capacity for social mobilization (Art. 24) & 
Deep synthesis service providers with public opinion attributes or capacity for social mobilization (Art. 19) & 
Generative AI service providers with public opinion attributes or capacity for social mobilization; required to carry out security assessments and formal filing (Art. 17) \\
\hline
\end{tabular}
\caption{Comparison of Algorithm-Related Regulatory Provisions in China}
\label{legalcomparison}
\end{table}

\section{Comparison of security self-assessment requirements across different algorithm types}

\begin{table}[H]
\centering
\small
\renewcommand{\arraystretch}{1.25}
\setlength{\tabcolsep}{13pt}
\begin{tabularx}{\columnwidth}{lccccccc}
\toprule
\textbf{Algorithm} &
\textbf{RM} &
\textbf{UR} &
\textbf{CE} &
\textbf{DE} &
\textbf{DS} &
\textbf{MS} &
\textbf{DSec} \\
\midrule
Personalized Recommendation      & ✔ & ✔ & ✔ & ✘ & ✔ & ✘ & ✘ \\
Information Retrieval            & ✔ & ✔ & ✔ & ✘ & ✔ & ✘ & ✘ \\
Content Filtering                & ✔ & ✔ & ✔ & ✘ & ✘ & ✘ & ✘ \\
Ranking / Selection              & ✔ & ✔ & ✔ & ✘ & ✔ & ✘ & ✘ \\
Scheduling / Dispatch             & ✔ & ✔ & ✘ & ✔ & ✘ & ✔ & ✘ \\
Generative AI (Tech Support)     & ✔ & ✔ & ✘ & ✘ & ✘ & ✔ & ✔ \\
Generative AI (Service Provider) & ✔ & ✔ & ✔ & ✘ & ✘ & ✔ & ✔ \\
\bottomrule
\end{tabularx}
\caption{Security self-assessment requirements by algorithm type. 
\ding{51}~indicates a required mechanism; \ding{55}~indicates not required. 
RM: Risk Mitigation; 
UR: User Rights Protection; 
CE: Content Ecosystem Governance; 
DE: Decision Evaluation; 
DS: Diversity Safeguards; 
MS: Model Security; 
DS: Data Security.}
\label{tab:self-assessment difference}
\end{table}

\section{Comparison of Algorithmic Service Beian and Large-Model Beian}

\begin{table}[H]
\centering
\renewcommand{\arraystretch}{1.3}
\begin{tabularx}{\textwidth}{|l|X|X|}
\hline
\textbf{Feature / Requirement} & \textbf{Algorithmic Service Beian} & \textbf{Large-Model Beian} \\
\hline
Scope (type of AI) & Personalized Recommendation, Information Retrieval, Content Filtering, Ranking / Selection, Scheduling-Related Decision-Making, Generative AI - Technical Support, Generative AI -  Service Provider & Large Generative Models \\
\hline
Regulatory Track & Standardized, fully online & Require offline engagement with local CAC \\
\hline
Legal Basis for Process & Clear, fully defined by CAC manual & Offline component lacks explicit legal basis; iterative consultation required \\
\hline
Core Information Required & 
\begin{enumerate}
    \item \textbf{Algorithm Entity Information}: business license, legal representative ID, algorithm responsible person ID and proof of employment, signed compliance letter, ICP filing/license
    \item \textbf{Algorithm Information}: basic attributes (name, category, application, purpose), detailed attributes (data sources, model architecture, algorithm logic, risk mitigation), and self-assessment
    \item \textbf{Product \& Function Information}: deployment context, access paths, functionality description, platform info, user exposure, personalization controls
\end{enumerate} & 
\begin{enumerate}
    \item \textbf{Algorithm Entity Information}: same as Algorithmic Service Beian
    \item \textbf{Algorithm Information}: extended self-assessment covering data and model security
    \item \textbf{Product \& Function Information}: same as Algorithmic Service Beian
    \item \textbf{Additional model-specific documents}:
    \begin{itemize}
        \item Model Service Agreement
        \item Data Annotation Rules
        \item Keyword Blocking List
        \item Evaluation Test Question Set
    \end{itemize}
\end{enumerate} \\
\hline
Review Approach & Procedural check via online system & Substantive review through iterative online + offline consultations, testing, and evaluation \\
\hline
\end{tabularx}
\caption{Comparison of Algorithmic Service Beian and Large-Model Beian, including details of Algorithm Entity and Algorithm Information requirements.}
\label{tab:Beian_comparison}
\end{table}

\section{Information Requirements for Tier 1 and Tier 2 ATRS Records}
\begin{table}[H]
\centering
\begin{tabular}{p{3.5cm} p{11cm}}
\hline
\textbf{Tier} & \textbf{Information Required} \\
\hline

\textbf{Tier 1 – Basic Information} &
High-level, concise information providing an overview of:
\begin{itemize}
    \item Name used to identify the algorithmic tool
    \item What the tool is
    \item Why it is being used
    \item URL for any related public-facing webpage
    \item Contact email for the team responsible for the tool
\end{itemize}
 \\

\hline

\textbf{Tier 2 – Detailed Information} &
\begin{minipage}[t]{0.48\linewidth}
\textbf{Owner and Responsibility}
\begin{itemize}
    \item Organisation or department
    \item Team
    \item Senior responsible owner
    \item Third-party involvement
    \item Companies House Number
    \item Third-party role
    \item Procurement procedure type
    \item Third-party data access terms
\end{itemize}

\textbf{Description and Rationale}
\begin{itemize}
    \item Detailed description
    \item Benefits
    \item Previous process
    \item Alternatives considered
\end{itemize}

\textbf{Deployment Context}
\begin{itemize}
    \item Integration into broader operational processes
    \item Human review
    \item Frequency and scale of usage
    \item Required training
    \item Appeals and review mechanisms
\end{itemize}
\end{minipage}
\hfill
\begin{minipage}[t]{0.48\linewidth}
\textbf{Tool Specification}
\begin{itemize}
    \item System architecture
    \item System-level inputs
    \item System-level outputs
    \item Maintenance
    \item Models
\end{itemize}

\textbf{Model Specification}
\begin{itemize}
    \item Model name
    \item Model version
    \item Model task
    \item Model input
    \item Model output
    \item Model architecture
    \item Model performance
    \item Datasets and their purposes
\end{itemize}

\textbf{Data Specifications}
\begin{itemize}
    \item Development data: description, modality, quantities, sensitive attributes, completeness, cleaning, collection, access, sharing
    \item Operational data: sources, sensitive attributes, processing, access, sharing
\end{itemize}

\textbf{Risks, Mitigations, and Impact Assessments}
\begin{itemize}
    \item Impact assessments
    \item Risks and mitigations
\end{itemize}
\end{minipage}
\\

\hline
\end{tabular}
\caption{Information Requirements for Tier 1 and Tier 2 ATRS Records}
\label{tab:ATRS_tiers_two_columns}
\end{table}

\section{Comparison of ATRS Information Delivery: Pilot Phase vs Current Implementation}
\begin{table}[H]
\centering
\begin{tabular}{p{2cm} p{5cm} p{5cm}}
\hline
\textbf{Aspect} & \textbf{Pilot Phase} & \textbf{Current Implementation} \\
\hline
\textbf{Information Tiers} & 
\textbf{Tier 1:} Proactively available at or before interacting with the algorithm; aimed at directly affected users. \newline
\textbf{Tier 2:} Available on demand for those seeking detailed information, typically intermediaries (experts, journalists, civil society). & 
Both tiers are published via a central online repository; Tier 1 remains concise and high-level, Tier 2 provides detailed, searchable information accessible to anyone. \\

\hline
\textbf{Delivery Channels} & 
Multiple channels suggested, including personalised or targeted communications to provide Tier 1 information. & 
Centralised online repository to simplify access, ensure consistency, and reduce administrative burden. \\

\hline
\end{tabular}
\caption{Comparison of ATRS Information Delivery: Pilot Phase vs Current Implementation}
\label{tab:ATRS_changes_comparison}
\end{table}

\section{Figures}

\begin{figure}[h]
\includegraphics[width=0.9\linewidth]{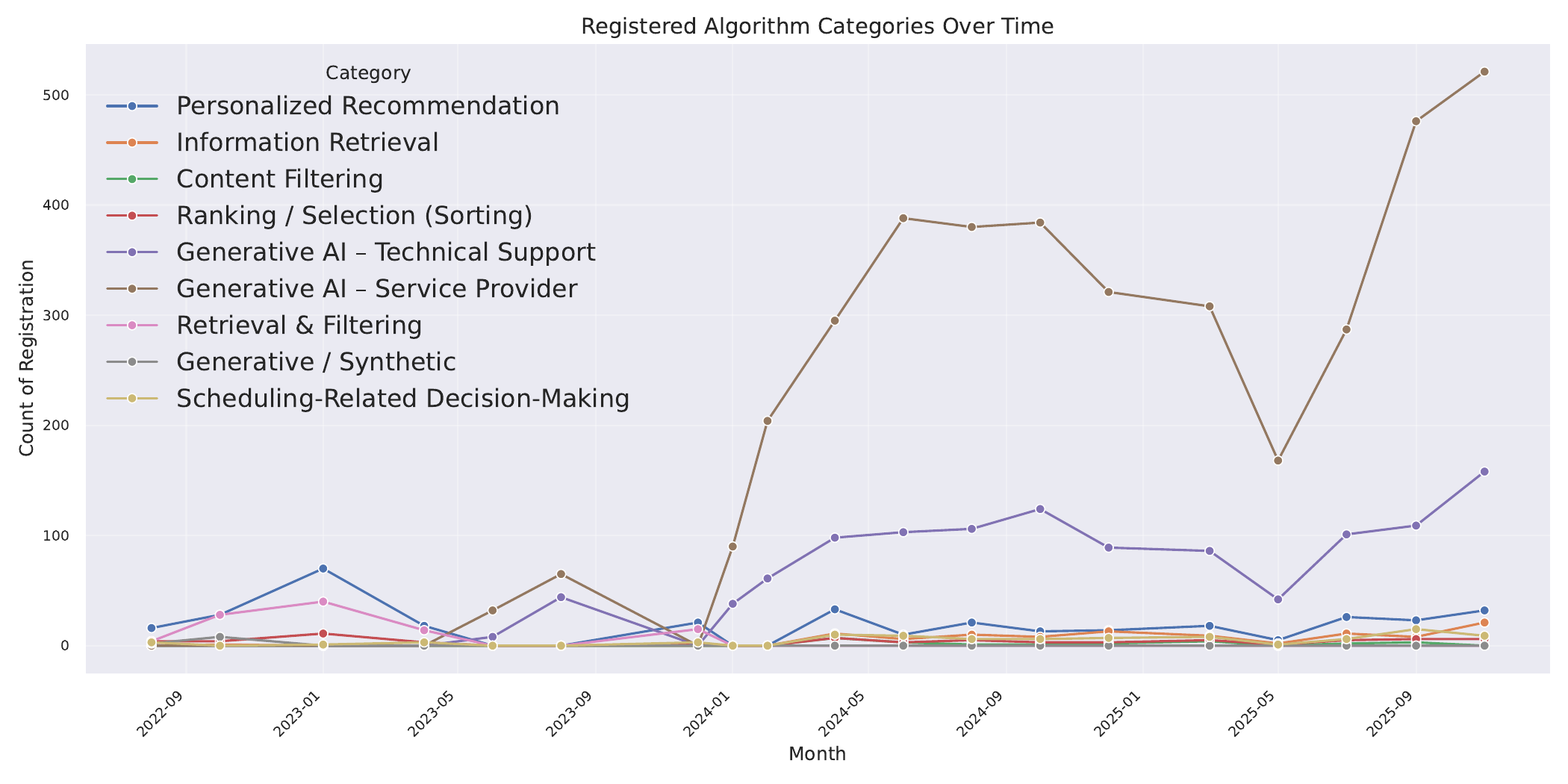}
    \caption{Algorithm registrations under Beian by category over time.}
    \label{fig:line_chart}
\end{figure}

\begin{figure}[h]
    \centering
    \includegraphics[width=0.9\linewidth]{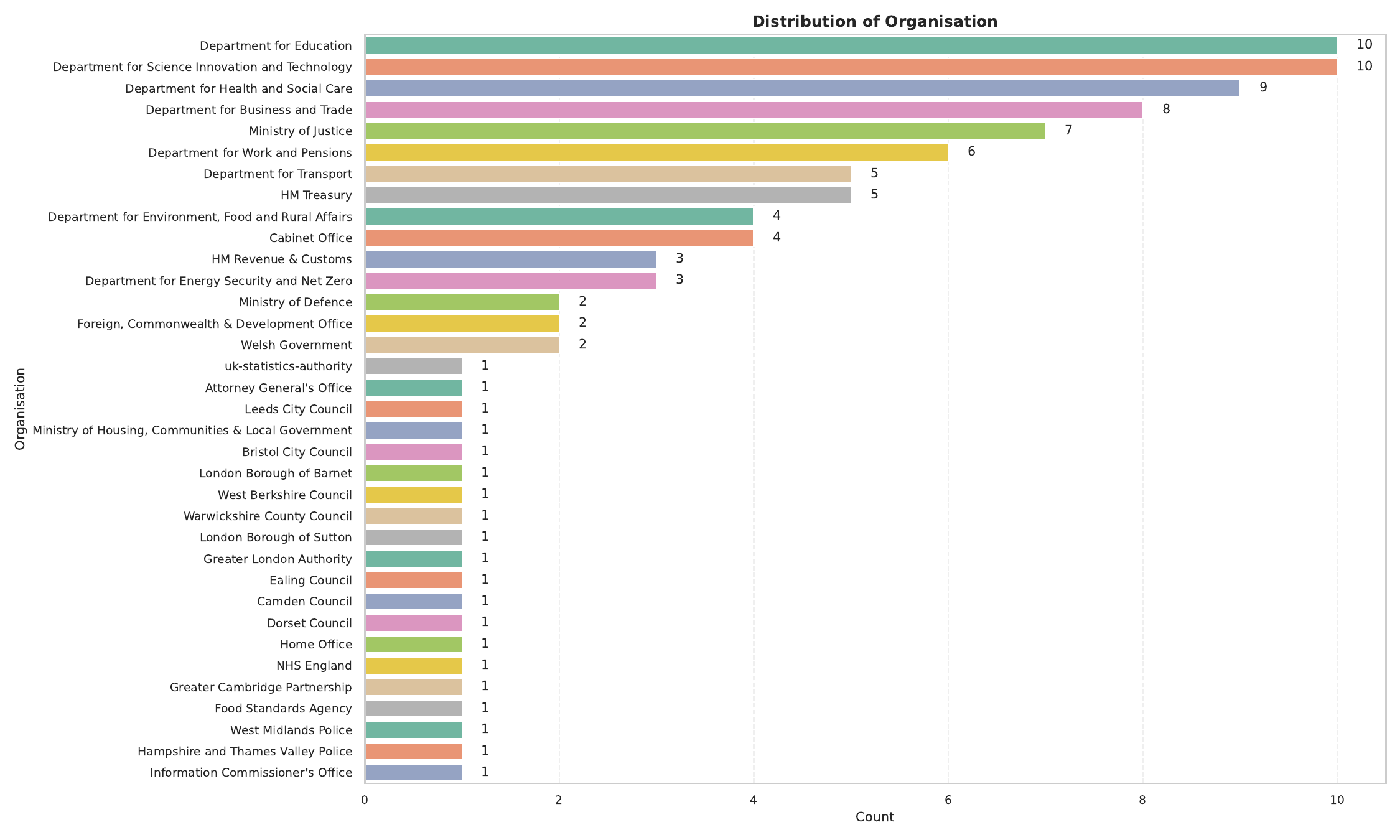}
    \caption{Registrations under ATRS by organization.}
    \label{fig:uk_register_by_department}
\end{figure}

\begin{figure}[h!]
    \centering
    \includegraphics[width=\linewidth]{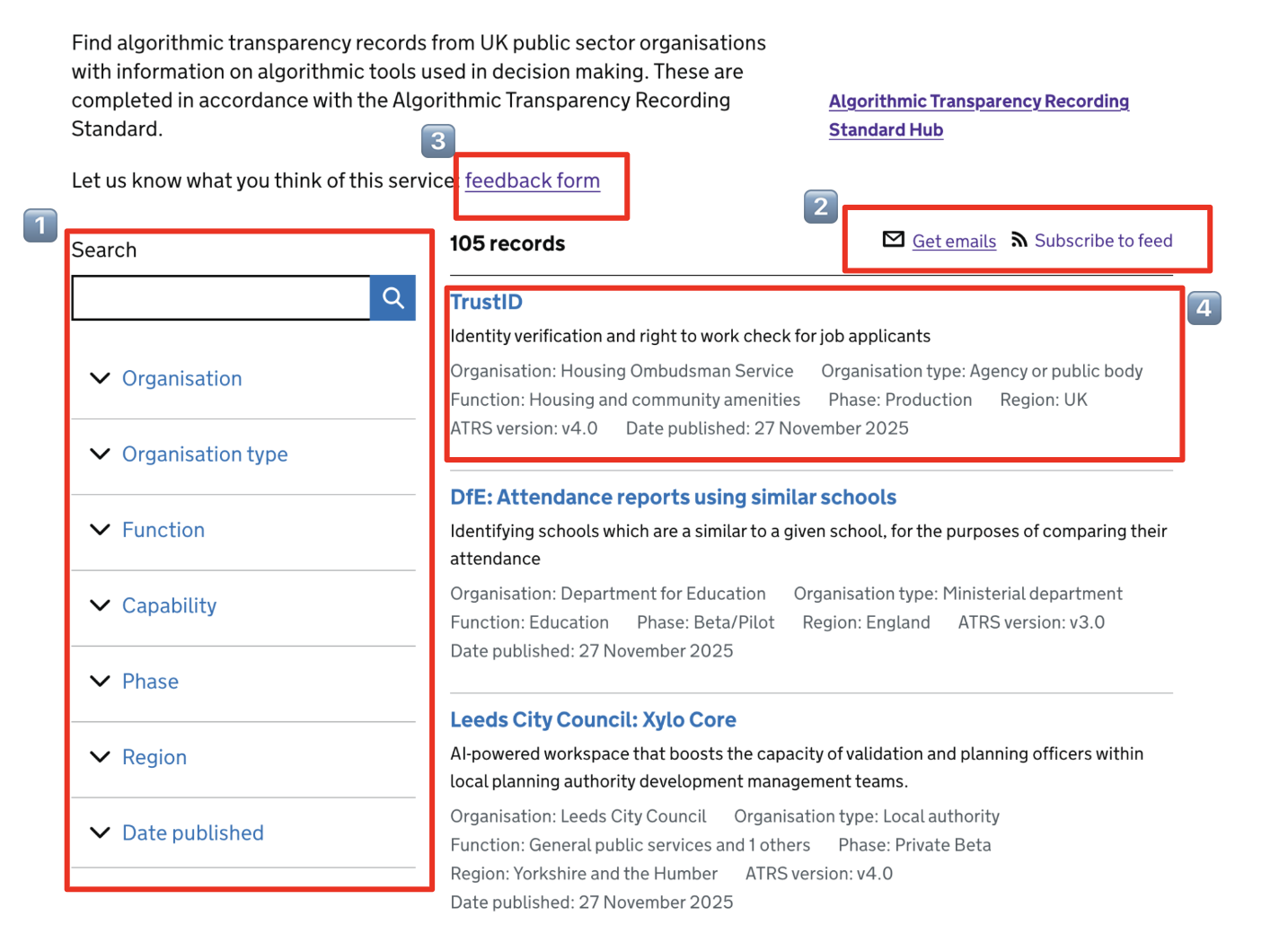} 
    \caption{Screenshot of the web-accessible interface captured on 12-15-2025. The interface provides: 
    (1) a search function supporting free-text queries and structured filters across organization, organization type, function, capability, deployment phase, region, and publication date; 
    (2) engagement mechanisms including ``Get emails'' and ``Subscribe to feed'' links that enable RSS-based monitoring; 
    (3) a feedback form for user input, suggestions, and issue reporting; and 
    (4) structured algorithm cards displaying the algorithm name, plain-language description, and key metadata.}
    \label{fig:web_interface}
\end{figure}

\end{document}